\documentclass[pra,twocolumn,floatfix]{revtex4}

\usepackage{graphicx} 
\usepackage{dcolumn}  
\usepackage{bm}       
\usepackage{amssymb}  
\usepackage{amsmath}  

\input epsf

\def\bea{\begin{eqnarray}}
\def\eea{\end{eqnarray}}
\def\ben{\begin{equation}}
\def\een{\end{equation}}
\def\benu{\begin{enumerate}}
\def\enu{\end{enumerate}}

\def\n{n}

\def\sss{\scriptscriptstyle\rm}

\def\g{_\gamma}

\def\1var{(\bx_1...\bx\N)}

\def\half{\frac{1}{2}}

\def\br{{\bf r}}

\def\bx{{\br t}}

\def\x{_{\sss X}}
\def\c{_{\sss C}}
\def\s{_{\sss S}}
\def\xc{_{\sss XC}}
\def\Hx{_{\sss HX}}

\def\Hxc{_{\sss HXC}}

\def\N{_{\sss N}}
\def\H{_{\sss H}}

\def\ext{_{\rm ext}}

\def\TF{^{\rm TF}}

\def\ee{_{\rm ee}}

\def\sph_int{ {\int d^3 r}}

\input rotate

\def\dr{d^{3}r}
\def\bei{\begin{itemize}}
\def\eei{\end{itemize}}
\def\z{_{\zeta}}
\def\zr{\zeta^{1/3}\br}

\def\alphau{^{\alpha}}
\def\ziu{^{\zeta}}
\def\giu{^{\gamma}}
\def\gr{\gamma \br}
\def\0{^{(0)}}

\def\ext{}
\def\NI{_{\sss NI}}
\def\NT{^{NT}}
\def\1s{_{\sss 1s}}
\def\2s{_{\sss 2s}}

\begin{document}

\title{Potential Scaling in Density Functional Theory}

\author{Peter Elliott}
\affiliation{Department of Physics and Astronomy, University of California, Irvine, CA 92697, USA}
\author{Kieron Burke}
\affiliation{Department of Chemistry, University of California, Irvine, CA 92697, USA \\ Department of Physics and Astronomy, University of California, Irvine, CA 92697, USA}


\begin{abstract}
Density scaling has a rich history in density functional theory,
providing exact conditions for use in the construction of ever
more accurate approximations to the unknown exchange-correlation functional.
We define a conjugate potential scaling for each density
scaling.  This provides exact relations on
various important density functionals, in particular, relating
functionals evaluated on exact densities of different potentials.
We illustrate these conditions on the two- and four-electron ion series.

\end{abstract}


\date{\today}

\maketitle


\section{Introduction}
Density functional theory (DFT) has become the method of choice for many electronic structure calculations in both computational chemistry and condensed-matter physics\cite{FNM03}. It balances the competing demands of accuracy and computational efficiency. The foundation of DFT is the Hohenberg-Kohn theorem\cite{HK64} which states that the total ground-state energy of a system can be written as a functional of the ground-state density. This allows the construction of the Kohn-Sham\cite{KS65} system of non-interacting fermions in a unique one-body potential, which is defined to yield the same density as the real interacting system. Since the ground-state energy is a density functional, this then gives the ground-state energy. However, this functional is not known exactly and in practice, the unknown exchange-correlation (XC) contribution must be approximated.

The plethora of XC functionals in current use in DFT is a symptom of 
the lack of a systematic method for functional 
development\cite{DCFB08}. In this regard, the more exact constraints that can be discovered, the more tools available for developers to use in their quest for more accurate approximations. Any useful approximation in DFT should try to satisfy these constraints, or be tested against them.

Density scaling has been a particularly useful tool for the analysis and development of DFT. A singular example is uniform coordinate scaling\cite{LP85}, where the coordinates of a given density are linearly scaled, but normalization is preserved. This has led to fundamental exact conditions on
the exchange-correlation (XC) energy functional\cite{LP85,L91,LP93,LO90}.
For example, the form of the local approximation to the exchange energy can be deduced from this scaling.
The adiabatic-connection formulation\cite{LP77,LP80,HJ74,GL76}, much studied and used in DFT development, is essentially an integral over the uniform coordinate scaling parameter\cite{LP85,Y87,LYP85}. Here, the electron-electron interaction is scaled by a constant while the density is kept fixed, linking the non-interacting Kohn-Sham and the fully interacting systems, and leads to many more conditions. For example, the adiabatic connection formula is behind rationalizing the hybrid approach\cite{PEB96,BEP97,BPEb97,E96}.

Recently, a different form of density scaling was used in the development of the PBEsol functional\cite{PRCV08}. Here, both the coordinate and the particle number are scaled, leading to new insights into the XC functional. We refer to this as charge-neutral scaling\cite{PCSB06}, as it is equivalent to simultaneously changing the charges on atoms and the number of electrons, so as to keep overall neutrality. 

In this paper, we extend the use of density scaling as a tool in DFT. Most importantly we introduce the concept that any form of density scaling defines a related form of potential scaling. This leads to more exact conditions on the various DFT quantities as functionals of densities of different particle number. Yang and others\cite{YAW04} have emphasized the duality of the potential with the density, but have not related scaling of one to the other.

\section{Background}
In DFT, the total ground-state energy for electrons in a given external potential, $v\ext(\br)$ is given as a functional of the ground-state density $n(\br)$,
\ben
E_{v\ext}[n] = F[n] + V\ext[n],
\een
where the external potential energy as a functional of the density is 
\ben
V\ext[n] = \int\dr \ n(\br)v\ext(\br),
\een
and $F[n]$ is the universal functional which may be defined via a constrainted search\cite{L79} over all wavefunctions
$\Psi$ that yield density $n(\br)$ 
\bea
F[n] &=& \min_{\Psi\rightarrow n} < \Psi |\hat{T} + \hat{V}\ee | \Psi>, \\
     &=& T[n] + V\ee[n] .
\eea
In the Kohn-Sham method, a reference system of non-interacting
electrons with the same density is solved, so it is useful to write
\ben
\label{Fdef}
F[n] = T\s[n] + U[n] + E\xc[n],
\een
where $T\s$ is the kinetic energy of non-interacting electrons of density 
$n(\br)$, $U$ is the Hartree energy and $E\xc$ is the 
exchange-correlation energy, defined by Eq. (\ref{Fdef}).
The Hartree energy (sometimes denoted by $E\H$) is defined as
\ben
\label{hart}
U[n] = \half\int\dr\int\dr' \frac{n(\br)n(\br')}{|\br -\br'|}.
\een
In Thomas-Fermi (TF) theory\cite{T26,F28}, the universal functional is approximated as 
\ben
F[n] \approx F\TF[n] = T\s^{(0)}[n] + U[n] \ ,
\een
where $T\s^{(0)}$ is the local kinetic energy functional,
\ben
T\s^{(0)}[n] =  A\s \int\dr \ n^{5/3}(\br) , 
\een
with $A\s = (3/10)(3\pi^2)^{2/3}$. 

\section{Potential Scaling}

Consider a density $n(\br)$ that is the ground-state density of some 
interacting problem with potential $v\ext(\br)$.  Now, introduce some positive parameter,
$0 < \gamma < \infty$, which produces a family of densities, $n\g(\br)$,
with $\gamma$ defined so that $\gamma\to\infty$ corresponds to the high-density
limit.
A simple example is the uniform coordinate scaling of Levy and Perdew\cite{LP85}: 
\ben
\label{uscal}
n\g(\br) = \gamma^3 n(\gr), ~~~~~0 < \gamma < \infty,
\een
where the prefactor was chosen to keep the density normalizd.
For example, under uniform coordinate scaling with $\gamma > 1$, the density of He is squeezed
into a smaller volume, and looks like a distorted version of a two-electron ion\cite{FTB00}. 
This scaling has become a mainstay of DFT and leads to many important results.
Most importantly, when particles interact,
the coordinate-scaled wavefunction is not the ground-state
wavefunction of the scaled density.  Considering such a wavefunction as
a trial state in the Rayleigh-Ritz principle yields
useful inequalities for the various density functionals\cite{LP85}: 
\ben
T[n\g] \leq \gamma^2\, T[n],~~~~~\gamma \geq 1,
\een
\ben
V\ee[n\g] \geq  \gamma\, V\ee[n],~~~~~~\gamma \geq 1,
\een
and a similar condition applies for the correlation energy $E\c[\n]$ itself.

\begin{figure}[htb]
\unitlength1cm
\begin{picture}(12,6)
\put(-6.45,-3.7){\makebox(12,6){
\includegraphics{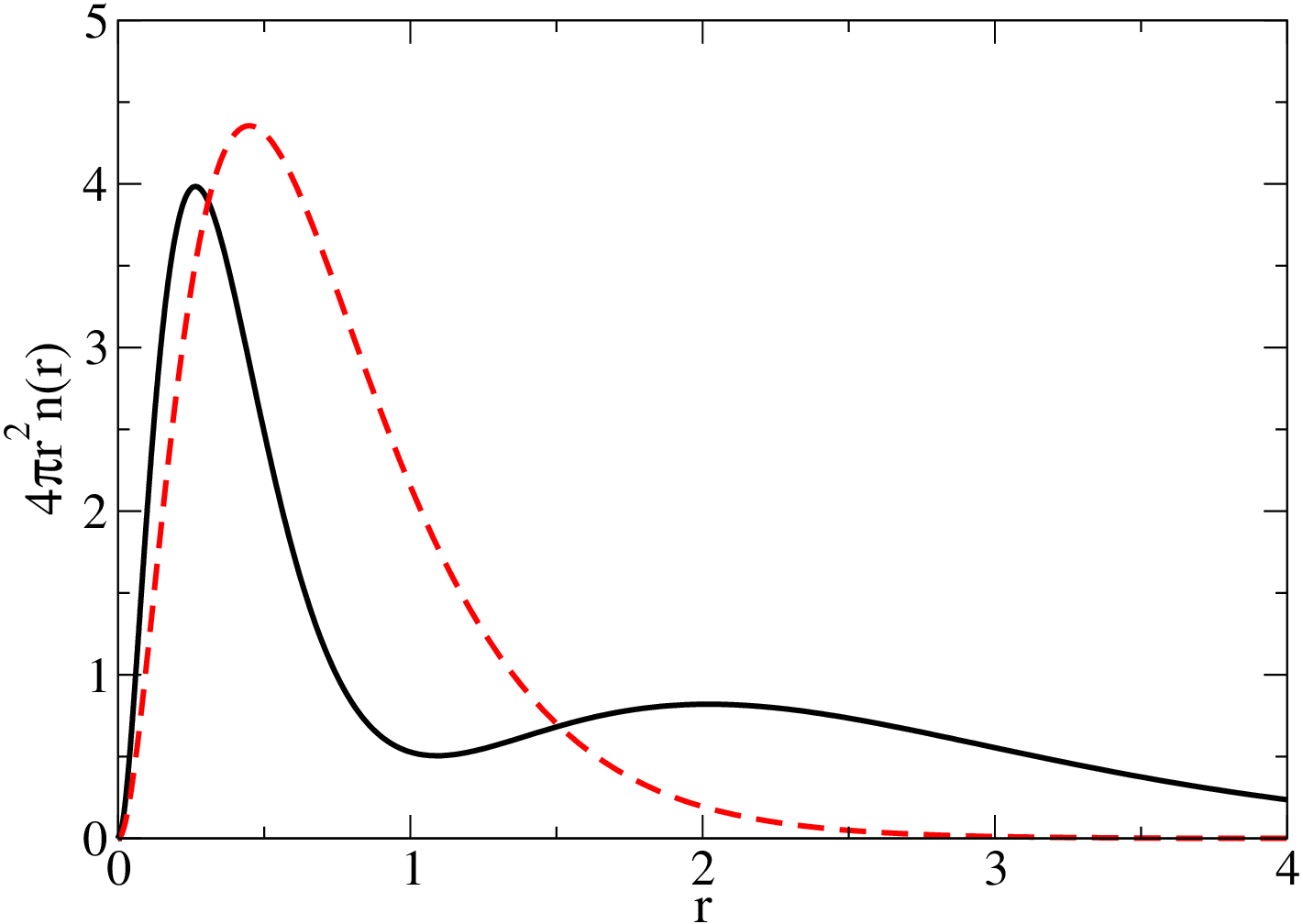}
}}
\end{picture}
\caption{The exact radial densities of Beryllium (solid line)\cite{UG93}, and of the CN scaled (with $\zeta=2$) Helium (dashed line)\cite{UG94}. }
\label{f:behescal}
\end{figure}

A second example that we focus on here is what we call charge-neutral (CN) scaling, 
in which 
\ben
\label{zscale}
n\z (\br) = \zeta^2 n(\zr), ~~~~~~~~0 < \zeta < \infty
\een
and so $N\z=\zeta N$. We use $\zeta$ as the scaling parameter to distinguish from coordinate scaling.  
This choice both scales the coordinate {\em and}
changes the particle number.  For Coulomb-interacting matter,
this ensures neutrality as a function of $\zeta$.  For example,
for single atoms, it simply implies $Z\z=\zeta Z$ and the atom remains
neutral. 
Lieb and Simon\cite{LS73} showed that Thomas-Fermi (TF) theory
becomes exact for neutral atoms as $\zeta\to\infty$, and Lieb\cite{L81}
later generalized the proof to all Coulomb-interacting matter.  In Fig 1, 
we illustrate this scaling on the He atom density.

In both coordinate and CN scaling, as the scaling parameter is taken to $\infty$,
the solution simplifies.  Under uniform coordinate scaling to 
the high-density limit, the system becomes effectively non-interacting.
Under CN scaling to the high-density limit,
Thomas-Fermi theory becomes relatively exact.  In either case, we can ask
how the potential changes when the density is scaled.  We define this
as the potential scaling conjugate to the given density scaling, but
consider it for all values of the scaling parameter, not just in the
high-density limit.

Under coordinate scaling, in the large $\gamma$ limit, 
\ben
\label{upotscal}
v\giu(\br) = \gamma^2v(\gr).
\een
We therefore define our potential scaling by this equation, applied for
all $\gamma$.
We use a superscript to indicate that the potential has been
scaled, not the density.
This is simply how the external potential would change when the density
is scaled, if the particles were {\rm non-interacting particles}.
For example, for  a neutral atom, this changes the nuclear charge by $\gamma$, keeping
the particle number fixed.  As $\gamma\to\infty$, the repulsion between
electrons becomes negligible relative to the nuclear attraction, and the
density becomes that of the non-interacting limit, scaled by $\gamma$.

Similarly, under CN scaling with $\zeta\to\infty$, 
the TF equations become relatively exact\cite{ELCB08}, and
\ben
\label{xiscal}
v\ziu(\br) = \zeta^{4/3}v(\zr),~~~~N_\zeta=\zeta N.
\een
Again, the conjugate potential scaling is defined by this, applied to
all values of $\zeta$.  Analogously, if self-consistent  TF
theory were exact, this is how the potential would scale for any
$\zeta$ as the density is scaled.

Although chosen to match the corresponding density scaling in the high-density
or high-potential limit, these potential scalings can be applied for any values
of their scaling parameter.  Since scaling the potential is much more common in
quantum problems than scaling the density, often solutions are
known or can be accurately calculated for different
scalings of the potential, but not of the density.  In this paper, we find
relations and inequalities between such solutions that complement the 
ground-breaking results of the previous generation\cite{LP85}.
\section{Uniform coordinate scaling}
In the old work\cite{LP85}, Levy and Perdew compared two different
wavefunctions with the same density, whereas we compare two different wavefunctions in the same potential. To do 
this, begin from a given potential $v\ext(\br)$ with ground-state density $\n(\br)$.  
Define  $n\giu(\br)$ as the ground-state density of $v\ext\giu(\br)$, given
by Eq. (\ref{upotscal}). 
Then $n\giu_{1/\gamma}(\br)$ is a useful trial density for the original problem.
It is found by first scaling the potential, solving the problem, and then scaling backwards to the original problem.  
(In Fig. 1, the dashed line corresponds $n\ziu(\br)$ for the He density, with $\zeta=2$.)
This is exactly what was done (but with an approximate scale factor) in Ref. \onlinecite{FTB00}.

If $n\giu_{1/\gamma}(\br)$ is used as trial density for $v\ext(\br)$, the variational principle states that
\ben
F[n\giu_{1/\gamma}] + \gamma^{-2}V\ext^{\gamma}[n\giu] 
\geq
 F[n] + \gamma^{-2}V\ext^{\gamma}[n\g],
\een
which may be rearranged as
\ben
F[n\giu_{1/\gamma}] - F[n] \geq \gamma^{-2}( V\ext^{\gamma}[n\g] - V\ext^{\gamma}[n\giu] ).
\een
Conversely, $n\g(\br)$ may be used as a trial density for $v\ext\giu(\br)$, yielding
\bea
E_{v\ext\giu}[n\g] &=& F[n\g] + V\ext\giu[n\g] \nonumber\\
			   &\geq& F[n\giu] + V\ext\giu[n\giu],
\eea
which can also be rearranged as
\ben
F[n\giu] - F[n\g] \leq V\ext\giu[n\g] - V\ext\giu[n\giu].
\een
Combining the two inequalities yields a constraint on the universal functional $F[\n]$:
\ben
F[n\giu_{1/\gamma}] - \frac{F[n\giu]}{\gamma^2} \geq F[n] - \frac{F[n\g]}{\gamma^2},
\een
which may be written in a concise form, with  $\lambda=1/\gamma$,
\ben
\label{Ineqgam}
\Delta F^\lambda[n_\lambda^{1/\lambda}] \geq \Delta F^\lambda[n],
\een
where
\ben
\label{delFg}
\Delta F^\lambda[n] = F[n] - \lambda^2 F[n_{1/\lambda}].
\een 
Now, $F[\n]$ is typically dominated by the kinetic energy contribution,
but this can be removed, because $T\s[\n\g]=\gamma^2\, T\s[\n]$.
Thus
\ben
\label{IneqHXC}
\Delta E\Hxc^\lambda[n_\lambda^{1/\lambda}] \geq \Delta E\Hxc^\lambda[n],
\een
where $E\Hxc=U+E\xc$.
This tells us that if we begin from, e.g., the lowest value of $Z$ that
binds a given $N$ electrons, then 
$\Delta E\Hxc^\lambda[n_\lambda^{1/\lambda}] $
is an increasing function of $\lambda$.

Simple results can be extracted from this very general formula by taking $\gamma$
to be very large.  This makes $\n^\gamma(\br)$ an essentially non-interacting
density, because the external potential dominates.  Thus
\ben
n_\lambda^{1/\lambda} (\br) \to n\NI (\br),~~~~~\lambda\to 0,
\een
where $n\NI(\br)$ is the density of the system with only an infinitesimal electron-electron
repulsion. But $\Delta E\Hxc^\lambda[\n]$ also simplifies as
$\lambda\to 0$, because all terms scale less than quadratically.
Thus
\ben
\Delta E\Hxc^\lambda[\n] \to  E\Hxc[\n],~~~~\lambda\to 0, 
\een
yielding the universal result that
\ben
E\Hxc[\n_{\sss NI}] \geq E\Hxc[\n],
\een
applying to all potentials. For $\gamma < 1$, Eq. (\ref{Ineqgam}) is less useful, as most systems
of interest lose an electron when the external potential becomes too small.  
To further simplify Eq. (\ref{IneqHXC}), we note that both the
Hartree and exchange energies scale linearly with $\gamma$, i.e.,
\ben
E\Hx[\n\g] = \gamma\, E\Hx[\n],
\een
so that
\ben
\Delta E\Hx^\lambda[\n] = (1-\lambda)\, E\Hx[\n].
\een
Inserted into Eq. (\ref{IneqHXC}), we find
\ben
\label{IneqHx}
E\Hx[\n^\gamma_\lambda]+\Delta' E^\lambda\c[\n^\gamma_\lambda] \geq
E\Hx[\n]+\Delta' E^\lambda\c[\n],
\een
where
\ben
\Delta' E^\lambda\c[\n] = \Delta E^\lambda\c[\n]/(1-\lambda).
\een
The simplest way to test this result is by doing a Kohn-Sham
calculation without any correlation (such as oep exact exchange).  Then the correlation
contributions vanish on both sides of  Eq. (\ref{IneqHx}), and
so
\ben
\label{EHXbounds}
E\Hx[n] \leq E\Hx[n\giu]/\gamma \leq E\Hx[\n\NI]
\een


\begin{table}
\caption{\label{tab:twoelec} The Hartree energies, $U$, for the helium iso-electronic series as calculated with the oep exact-exchange method as implemented in the OPMKS code\cite{opmks}. We also demonstrate how, for two values of atomic number $Z'$, the inequalities of Eq. (\ref{Uscale}) with $\gamma=Z'/Z$, are satisfied. Note that if $\gamma<1$, the inequality is reversed. The values for bordering values of Z bracket the value of $U$ at atomic number $Z$ and these bounds become tighter as $Z'$ increases.}
\begin{ruledtabular}
\begin{tabular}{c|ccc}
Z		&U			& Z'=4		& Z'=20		\\
\hline
1		& 0.790970	& 3.163880	& 15.819400 \\
2		& 2.051538	& 4.103076	& 20.515380 \\
3		& 3.303373	& 4.404497	& 22.022487 \\
4		& 4.554137	& 4.554137	& 22.770685 \\
6		& 7.054819	& 4.703213	& 23.516063 \\ 
10		& 12.055315	& 4.822126	& 24.110630 \\
20		& 24.555661	& 4.911132	& 24.555661 \\
\end{tabular}
\end{ruledtabular}
\end{table}

\begin{table}
\caption{\label{tab:fourelec} Hartree-exchange energies for the beryllium iso-electronic series. Values were also calculated with the OPMKS code with oep exact-exchange. Also shown are two examples of the inequalities of Eq. (\ref{EHXbounds}), again using $\gamma=Z'/Z$. Although the quantities are more complicate that those in Table \ref{tab:twoelec}, the overall trend is the same.}
\begin{ruledtabular}
\begin{tabular}{cc|ccc}
Ion			& Z			& E$\Hx$	& Z'=10		& Z'=16		\\
\hline
Be			& 4			& 4.489776	& 11.224440	& 17.959104	\\
B$^{+}$		& 5			& 6.119120	& 12.238240	& 19.581184	\\
O$^{4+}$	& 8			& 10.893545	& 13.616931	& 21.787090	\\
Ne$^{6+}$	& 10		& 14.051482	& 14.051482	& 22.482371	\\
S$^{12+}$	& 16		& 23.498356	& 14.686473	& 23.498356	\\
Ca$^{16+}$	& 20		& 29.788628	& 14.894314	& 23.830902	\\

\end{tabular}
\end{ruledtabular}
\end{table}


\begin{figure}[htb]
\unitlength1cm
\begin{picture}(12,7)
\put(-6.45,-3.7){\makebox(12,7){
\includegraphics{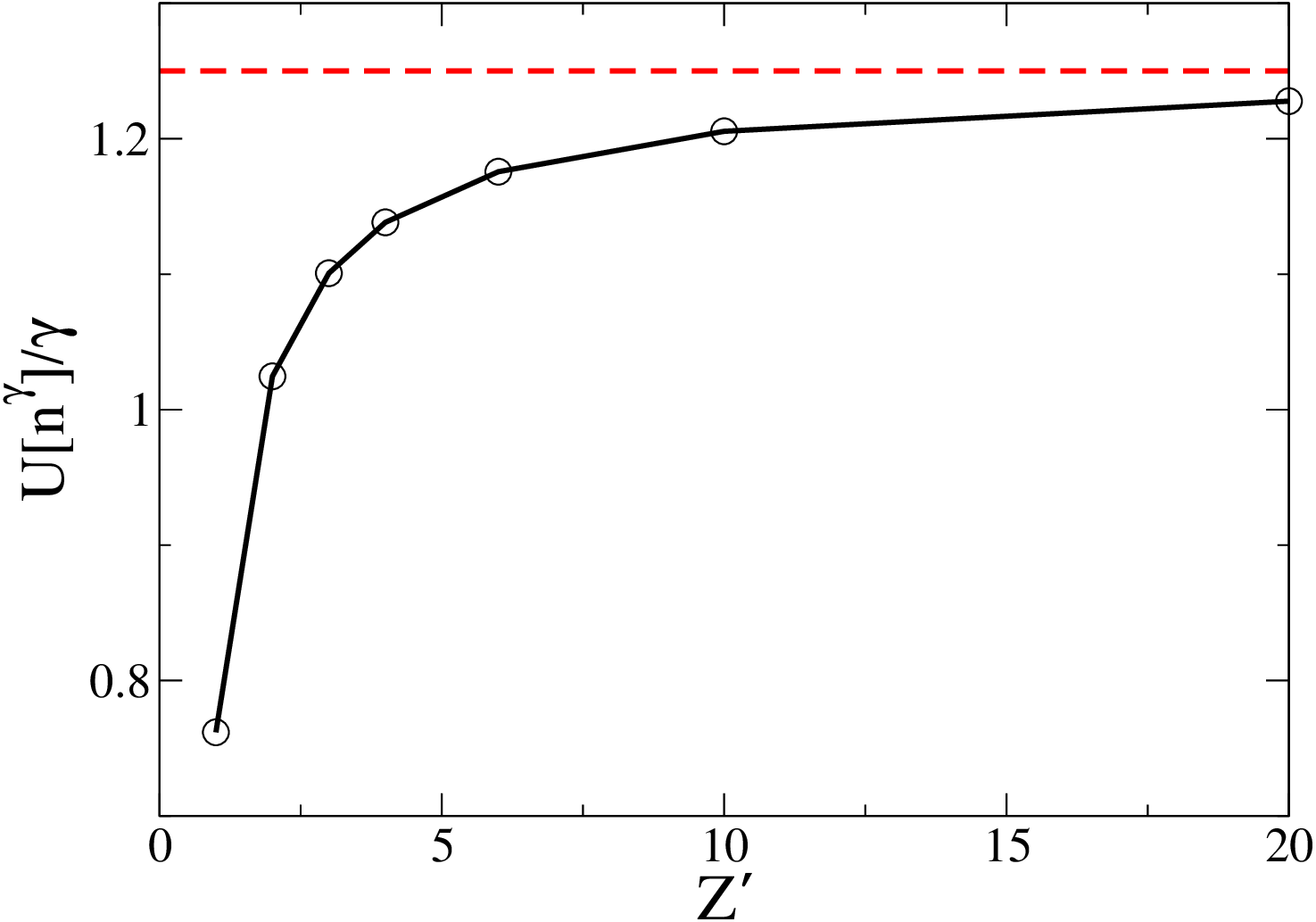}
}}
\end{picture}
\caption{Using the Hartree energies from Table \ref{tab:twoelec}, Eq. (\ref{Uscale}) is illustrated for $\gamma=Z'/Z$ and $Z=1$. The trend is identical to that seen in Table \ref{tab:twoelec}, however it is clear that the value is approaching it's asymptote, $5/4$. This is the Hartree energy for density consisting of the doubly occupied hydrogen $1s$ orbital.}
\label{f:UHion}
\end{figure}

\begin{figure}[htb]
\unitlength1cm
\begin{picture}(12,7)
\put(-6.45,-3.7){\makebox(12,7){
\includegraphics{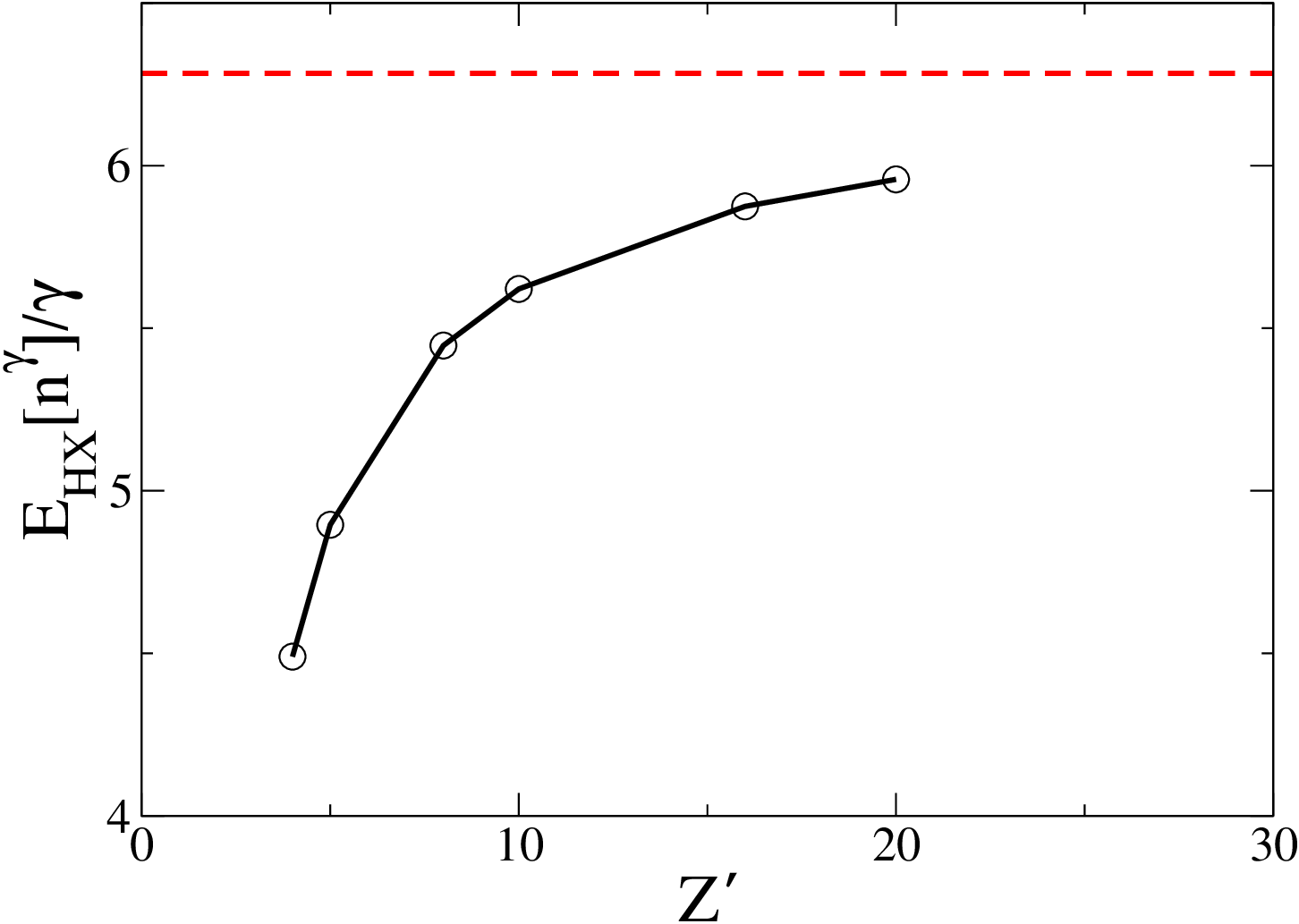}
}}
\end{picture}
\caption{The Hartree-exchange energies reported in Table \ref{tab:fourelec} are used to illustrate the inequalities of Eq. (\ref{EHXbounds}) with $\gamma=Z'/Z$ and $Z=4$. Compared to Fig. \ref{f:UHion}, the value of $E\Hx[n\giu]/\gamma$ is not as fully converged to its asymptote, however the maximum value of $\gamma$ is $4$ times smaller. The asymptotic value for this case is $586373/93312 = 6.284$, which is found by doubly occupying both 1s and 2s hydrogenic orbitals with $Z=4$ and calculating Hartree and exchange energies.}
\label{f:EHXBE}
\end{figure}

This simplifies even further for the special case of two electrons in a spin
singlet, where
$E\x[\n] = -U[\n]/2$, so the inequality becomes a bound on the Hartree energy
alone:
\ben
\label{Uscale}
U[n] \leq U[n\giu]/\gamma \leq U[\n\NI]
\een
In Table \ref{tab:twoelec}, we analyze the above inequality, Eq. (\ref{Uscale}), while in Fig \ref{f:UHion}, we plot $U[\n\giu]/\gamma$ as a function of $\gamma$ for exact-exchange calculations of the two-electron ion series, beginning with H$^-$. Indeed, the function increases
toward the Bohr atom limit of 5/4, found by inserting a doubly-occupied 1s Hydrogen
atom orbital into the Hartree energy.

To test the exchange contribution in a non-trivial way, i.e., Eq. (\ref{IneqHx}),
we repeated the calculations for the four-electron ion series, this time beginning
from Be.  Again the inequality is satisfied, and the limiting value is found
by evaluating the Hartree and exchange energies of
doubly-occupied 1s and 2s Hydrogenic orbitals, as calculated in Appendix \ref{AExNIBe}. These values are reported in Table \ref{tab:fourelec} and plotted in Fig \ref{f:EHXBE}.

Lastly, we can even include extremely accurate estimates of the correlation
contributions for the two-electron series.  We work from the data in 
Table I of Ref. \cite{HU97}.   Since the two-electron ions are generally
weakly correlated, one can approximate the scaling of their correlation
energies with a Taylor-series around the high-density limit:
\ben
E\c[\n] = E\c^{(0)}[\n] + \lambda\, E\c^{(1)} [\n]
\een
where $E\c^{(p)} [\n]$ are scale-invariant functionals.  Since $T\c=-E\c+ \partial
E\c[\n\gamma]/\partial \gamma (\gamma=1)$\cite{LP85}, and $T\c$ is reported in
their table, one can solve for these two coefficients.  This yields a value
of -47.6 mH for $E\c^{(0)}$ for He, in excellent agreement with the value
of 47.9 estimated in Ref. \cite{FTB00}, and predicts a value of -56.1 mH for
H$^-$.  Using this approximate scaling, we can insert all terms into Eq. (\ref{IneqHXC})
explicitly and find their behavior.  The numerical corrections to our
previous results are negligible.


\section{Charge-neutral Scaling}

In this section, we repeat all the logic of the previous section, but
apply it now to CN scaling. After repeating similar steps (given in Appendix \ref{ATFscal}), we arrive at the general result:
\ben
\label{ineqTF}
\Delta F\alphau[n^{1/\alpha}_{\alpha}] \geq \Delta F\alphau[n] \ ,
\een
where
\ben
\Delta F\alphau[n] = F[n] - \alpha^{7/3}F[n_{1/\alpha}] \ ,
\een
and $\alpha=1/\zeta$. Just as we did for coordinate
scaling, we can refine our inequality substantially. By construction, $\Delta F\alphau[n]=0$ for $F\TF[\n]$, so
we define the useful functional:
\ben
F\NT[\n] = F[\n] -F\TF[\n] 
\een
as the {\em Non}-Thomas-Fermi contribution to $F[\n]$. Our inequality then reads:
\ben
\label{ineqNTF}
\Delta F^{NT\alpha}[n^{1/\alpha}_{\alpha}] \geq \Delta F^{NT\alpha}[n] \ ,
\een
where
\ben
\Delta F^{NT\alpha}[n] = F\NT[n] - \alpha^{7/3}F\NT[n_{1/\alpha}] \ .
\een
We find an interesting result in the
limit $\alpha\to 0$, if we make the reasonable assumption that
all non-Thomas-Fermi contributions scale less strongly than $\zeta^{7/3}$ :
\ben
F^{NT}[n\TF] \ge F^{NT}[n] \ ,
\een
as TF becomes relatively exact in the high $\zeta$. 
This inequality is fiendishly hard to test, even in the large $\zeta$ limit.
Consider, e.g., the He atom.  The corresponding TF density is well-known\cite{LCPB09}
but we would have to evaluate the exact interacting functional on it to find the non-TF contribution.
All the above results also apply directly to non-interacting electrons
in a potential, such as the Bohr atom\cite{E88}, with $F$ replaced by $T\s$, and the
TF contributions calculated with no Hartree term.  But the same difficulties
remain.

There is one case where we know enough already to test.  For the hydrogen
atom (or any one-electron system), $F=T$ only, and is given by the von Weizacker
functional.  The TF density (with or without interaction) is well-known and singular
at the origin, making the von Weizacker energy diverge.  Thus, the formula
is satisfied, but not very informative.

Lastly, we consider
 Thomas-Fermi-Dirac-Weizs\"{a}ker theory\cite{Kc57} (TFDW).
Here we add to TF the local exchange 
\ben
E\0\x[n] = A\x \int\dr \ n^{4/3}(\br),
\een
where $A\x = -(3/4)(3/\pi)^{1/3}$, 
and the next order gradient correction
to the kinetic energy,
\ben
T\s^{(2)}[n] = \frac{1}{72}\int\dr \ \frac{|\nabla n(\br)|^2}{n(\br)}.
\een
Both these terms scale the same way under CN density scaling, i.e.,
\ben
F^{(2)}[n\z] = T^{(2)}[n\z] + E^{(0)}\x[n\z] = \zeta^{5/3}( T^{(2)}[n] + E^{(0)}\x[n] ).
\een
Then can write the inequality as
\ben
F^{(2)}[n\ziu] \geq \zeta^{5/3} F^{(2)}[n],
\een
where $\n(\br)$ has been evaluated self-consistently within TFDW and $\zeta \ge 1$.
Thus
\ben
F^{(2)}[n\TF] \geq F^{(2)}[n\ziu]/\zeta^{5/3} \geq  F^{(2)}[n],
\een
where $n\TF(\br)$ is the Thomas-Fermi solution for the same potential as for $n(\br)$.

\section{Conclusion}

Potential scaling, conjugate to a given density scaling, promises to be a useful tool
in density functional theory. It leads to many exact conditions that can be used in functional construction.
We have applied it to two distinct types of scaling:  uniform coordinate scaling and charge neutral scaling. In both cases, we have found several interesting bounds.
Uniform coordinate scaling was useful for analyzing Kohn-Sham DFT, leading to inequalities involving the only unknown in DFT, the exchange-correlation functional. The limit of this inequality involves evaluating the Hartree-exchange-correlation energy of the density of non-interacting fermions in the external potential. This connection between the interacting and non-interacting systems resonates with standard approaches in many-body perturbation theory. We illustrate the bounds on the Hartree-exchange energy this inequality provides by performing OEP exact exchange calculations on helium and beryllium, showing the approach to their asymptote.
On the other hand, charge-neutral scaling provides inequalities involving Thomas-Fermi quantities. The Thomas-Fermi approximation becomes relatively exact for all electronic systems\cite{LS73,L81} and these relations link the corrections to Thomas-Fermi with the true system, including those of TFDW theory. However evaluating the TF density within these theories often leads to divergences\cite{M81,YPKZ97}.
It is not clear how important these relations will be for functional development, but those derived from uniform coordinate scaling are siblings to those derived by Levy and Perdew\cite{LP85}, which proved very useful in constraining approximations in DFT. 

We thank Eberhard Engel for use of the OPMKS code and also thank Cyrus Umrigar for providing exact densities. We acknowledge support from NSF under grant CHE-0809859.

\appendix
\section{Charge-neutral scaling inequality}
\label{ATFscal}
We follow the steps in deriving Eq. (\ref{Ineqgam}) but applied to the charge neutral scaling defined in Eq. (\ref{xiscal}). Taking $n\ziu_{1/\zeta}(\br)$ as a trial density for the $v(\br)$ system, then the variational principle states:
\ben
\label{In1Ev}
F[n\ziu_{1/\zeta}] + V[n\ziu_{1/\zeta}] \geq F[n] + V[n] \ ,
\een
where $\n\ziu(\br)$ is the exact density for the scaled potential $v\ziu(\br)$. Conversely, use $n\z$(\br) as a trial density for the $v\ziu(\br)$ system:
\ben
F[n\z] + \zeta^{7/3}V[n] \geq  F[n\ziu] + \zeta^{7/3}V[n\ziu_{1/\zeta}]
\een
where we have used $V\ziu[n\z] = \zeta^{7/3}V[n]$. Combining these inequalities gives:
\ben
\label{In1F}
\frac{F[n\z] - F[n\ziu]}{\zeta^{7/3}} \geq F[n] - F[n\ziu_{1/\zeta}] \ ,
\een
which may be written as 
\ben
\Delta F\ziu[n\ziu_{1/\zeta}] \geq \Delta F\ziu[n] \ ,
\een
with
\ben
\label{AineqTF}
\Delta F\ziu[n] = F[n] - \frac{F[n\z]}{\zeta^{7/3}} \ .
\een

\section{Exchange energy for non-interacting Beryllium}
\label{AExNIBe}
The limit of the inequality, Eq. (\ref{IneqHx}), is the Hartree-exchange functional evaluated on the density of the corresponding non-interacting system. Since the g.s. orbitals which sum to this density are known analytically (they are simply hydrogenic orbitals), we may calculate the exact Hartree-exchange value.
 
Written in spherical coordinates, $ r = |\br| $, the $1s$ and $2s$ hydrogenic orbitals are:
\bea
\phi_{\sss 1s}(\br) &=&\left( \frac{Z^3}{\pi}\right)^{1/2}e^{-Zr} \ ,\\
\phi_{\sss 2s}(\br) &=& \left( \frac{Z^3}{32\pi}\right)^{1/2}(2 - Zr) e^{-Zr/2} \ .
\eea
For beryllium, both these orbitals are doubly occupied, giving the total density as
\ben
n(\br) = 2|\phi_{\sss 1s}(\br)|^2 + 2|\phi_{\sss 2s}(\br)|^2 \ .
\een
The Hartree energy is defined by Eq. (\ref{hart}), however in the special case of spherical densities it may be written as:
\ben
\label{Usph}
U[n] = \half \int_0^\infty dr ~ \left(f[n](r)\right)^2 \ ,
\een
where 
\ben
\label{fsph}
f[n](r) = 4\pi\int_r^\infty dr' ~ r' n(r') \ ,
\een
and we use square brackets to indicate that it is a functional of the density. The exchange energy for a spin-unpolarized system is:
\ben
\label{Ex}
E\x = -2 \half \sum_{i,j}^{occ}\sph_int \int d^3 r' ~ \frac{\phi^*_i(\br)\phi^*_j(\br')\phi_j(\br)\phi_i(\br')}{|\br-\br'|} \ ,
\een
where the factor $2$ is due to spin, and the sum is over occupied orbitals only, in this case $1s$ and $2s$. If we define a new quantity, $\tilde{n}(\br)$: 
\ben
\tilde{n}(\br) = \phi_{\sss 1s}(\br)\phi_{\sss 2s}(\br) \ ,
\een
then we may write
\ben
E\x = -2\left( U[n\1s] + 2U[\tilde{n}] +U[n\2s] \right) \ .
\een
We can use Eq. (\ref{fsph}) and Eq. (\ref{Usph}) for each term separately and then combine to find the total exchange energy. The answer will be equivalent to solving Eq. (\ref{Ex}) using the orbitals, however this method avoids performing integrals involving $1/|\br-\br'|$ and, in this case, are easy to solve using integration by parts.
The values for Hartree, exchange and their sum are:
\[ U[n] = \frac{5 \times 23 \times 431}{2^6 3^4} = \frac{49565}{5184} = 9.561 \ , \]

\[ E\x = -\frac{59 \times 71 \times 73}{2^7 3^6} = -\frac{305797}{93312} = -3.277 \ , \]

\[ E\Hx = \frac{383 \times 1531}{2^7 3^6} = \frac{586373}{93312}= 6.284 \ . \]

\end{document}